\documentclass[nofootinbib,twocolumn]{revtex4-2}
\usepackage[T1]{fontenc}
\usepackage{graphicx,subcaption}
\usepackage{rotating}
\usepackage{bm}        
\usepackage{amssymb}   
\usepackage{float}
\usepackage{tikz}
\usepackage{diagbox}
\usepackage{braket}
\usepackage{color} 
\usepackage{colortbl}
\usepackage{amsmath}
\usepackage{xcolor}
\usepackage{soul}
\usepackage{float,subcaption}
\usepackage[rightcaption]{sidecap}
\DeclareCaptionJustification{justified}{\justifying}
\usepackage{booktabs}
\usepackage{siunitx}
\usepackage{caption}
\usepackage{multirow}
\usepackage{array}
\usepackage{chemfig}
\setchemfig{atom sep=2em}
\usepackage{tabularx}
\newcommand{\req}[1]{\begin{align}#1\end{align}}

\makeatletter

    \def\CT@@do@color{%
      \global\let\CT@do@color\relax
            \@tempdima\wd\z@
            \advance\@tempdima\@tempdimb
            \advance\@tempdima\@tempdimc
    \advance\@tempdimb\tabcolsep
    \advance\@tempdimc\tabcolsep
    \advance\@tempdima2\tabcolsep
            \kern-\@tempdimb
            \leaders\vrule
                    \hskip\@tempdima\@plus  1fill
            \kern-\@tempdimc
            \hskip-\wd\z@ \@plus -1fill }
    \makeatother

\def\k1{k_1}
\def\k2{k_2}
\def\q1{q_1}
\def\q2{q_2}

\def\({\left (}
\def\){\right )}
\def\[{\left [}
\def\]{\right ]}

\newcommand{\beq}{\begin{equation}}
\newcommand{\eeq}{\end{equation}}

\DeclareMathAlphabet\mathbfcal{OMS}{cmsy}{b}{n}

\begin{document}

\title{Molecular representations of quantum circuits for quantum machine learning}
\author{Elham Torabian and Roman V. Krems}
 \affiliation{
Department of Chemistry, University of British Columbia, Vancouver, B.C. V6T 1Z1, Canada \\
Stewart Blusson Quantum Matter Institute, Vancouver, B.C. V6T 1Z4, Canada }

\date{\today}

\begin{abstract}

We establish an isomorphism between quantum circuits and a subspace of polyatomic molecules,
which suggests that molecules can be used as descriptors of quantum circuits for quantum machine learning. 
Our numerical results show that the performance of quantum circuits for quantum support vector machines can be characterized by dimensionality-reduced molecular fingerprints as well as by the size of the largest and smallest Gershgorin circles derived from the Coulomb matrices of the corresponding molecules. This can be used to restrict the search space for the compositional optimization of quantum circuits. We show that a high accuracy of a quantum algorithm can be achieved with high probability by sampling from a specific set of molecules. 
This work implies that quantum ansatz optimization can benefit from advances in cheminformatics and suggests an approach to identify key elements that enhance the accuracy of a quantum algorithm by mapping quantum circuits onto molecules and exploring correlations between physical properties of molecules and circuit performance.

\end{abstract}

\maketitle

\section{Introduction}

Recent evidence suggests that general quantum learning algorithms are either efficiently classically simulable, or suffer from limitations due to vanishing of inner products of quantum states in exponentially large Hilbert spaces \cite{kubler2021inductive}, which leads to barren plateaus (BP) in variational algorithms \cite{mcclean2018barren,larocca2024review, zhang2024absence, sack2022avoiding, holmes2021barren, friedrich2022avoiding, kulshrestha2022beinit} or exponential concentration (EC) of quantum kernels \cite{arrasmith2022equivalence,thanasilp2024exponential, kairon2025equivalence}.
BP and EC thus appear to preclude the quantum advantage of quantum machine learning (QML). 
On the other hand, it has been shown that quantum kernels can solve classification problems with NP \cite{liu2021rigorous} and PromiseBQP-complete complexity \cite{jager2023universal}, indicating that the quantum advantage of QML can be achieved, in principle. These results suggest that,  for practical applications, the search for the quantum advantage of QML should focus on general strategies of encoding inductive bias
into the quantum algorithms (see also \cite{kubler2021inductive}). 

For gate-based quantum computing, quantum states are generated by circuits of logic gates consisting of unitary operators acting on a given initial state of qubits. 
The most general strategy for constructing optimally biased quantum circuits (QC) is the optimization of operator sequences in the space of gate permutations. 
However, the computational complexity of such optimization scales exponentially with the number of gates and qubits, even when greedy algorithms are employed \cite{torabian2023compositional}.  
To reduce this complexity, it is necessary to restrict the search space. This is usually achieved by either assuming a specific structure for the quantum ansatz or by building adaptive-structure ansatze from simple quantum circuits.  QC can also be chosen to respect relevant symmetries.  
For example, EC can be avoided by restricting the space of QC to covariant quantum kernels \cite{glick2024covariant}, or by quantum kernel bandwidth methods \cite{shaydulin2022importance}, quantum Fisher kernels \cite{suzuki2022quantum}, and projected quantum kernel methods \cite{huang2021power}. However, these methods rely on some prior information about the data structure. 

In the present work, we aim to restrict the search space of gate permutations, extending to deep circuits, by parameters determined from optimization of shallow circuits. 
More specifically, we consider the following problem: given a set of QC with depth ${\cal P} < {\cal T}$, identify the predictors of the performance of  QC with depth ${\cal P} > {\cal T}$. To achieve this, we require 
an algorithm for QC construction that satisfies the following conditions: (i)  deep QC must inherit relevant properties of shallow circuits; (ii)  the performance of QC of any depth can be discriminated by efficiently (polynomial in the number of gates and qubits) computable metrics.   
To develop an algorithm that meets these conditions, we establish and exploit an isomorphism between QC and polyatomic molecules. 
With a carbon polymer backbone as the qubit structure,  we propose a mapping that can be exploited to use molecules as descriptors of quantum circuits. 
We then demonstrate that the shape of the resulting molecules, encoded in the Gershgoring circles of Coulomb matrices for the spatial atomic arrangements, yields predictors of QC performance. 
With the proposed circuit - molecule mapping, the computation of the Gershgoring circles is, at most, quadratic in the number of qubits and gates. Our results demonstrate that these conclusions generalize both to different samples within a given data set and to different datasets. 
Our results imply that mapping QC onto physical objects, such as molecules, can be used to elucidate the important features of the quantum ansatz design.  
Our results also suggest that methods developed for optimization of molecular properties in chemical compound spaces for drug-design applications, can be transferred to QC optimization.

\begin{figure*}
    \centering
    \includegraphics[scale=0.6]{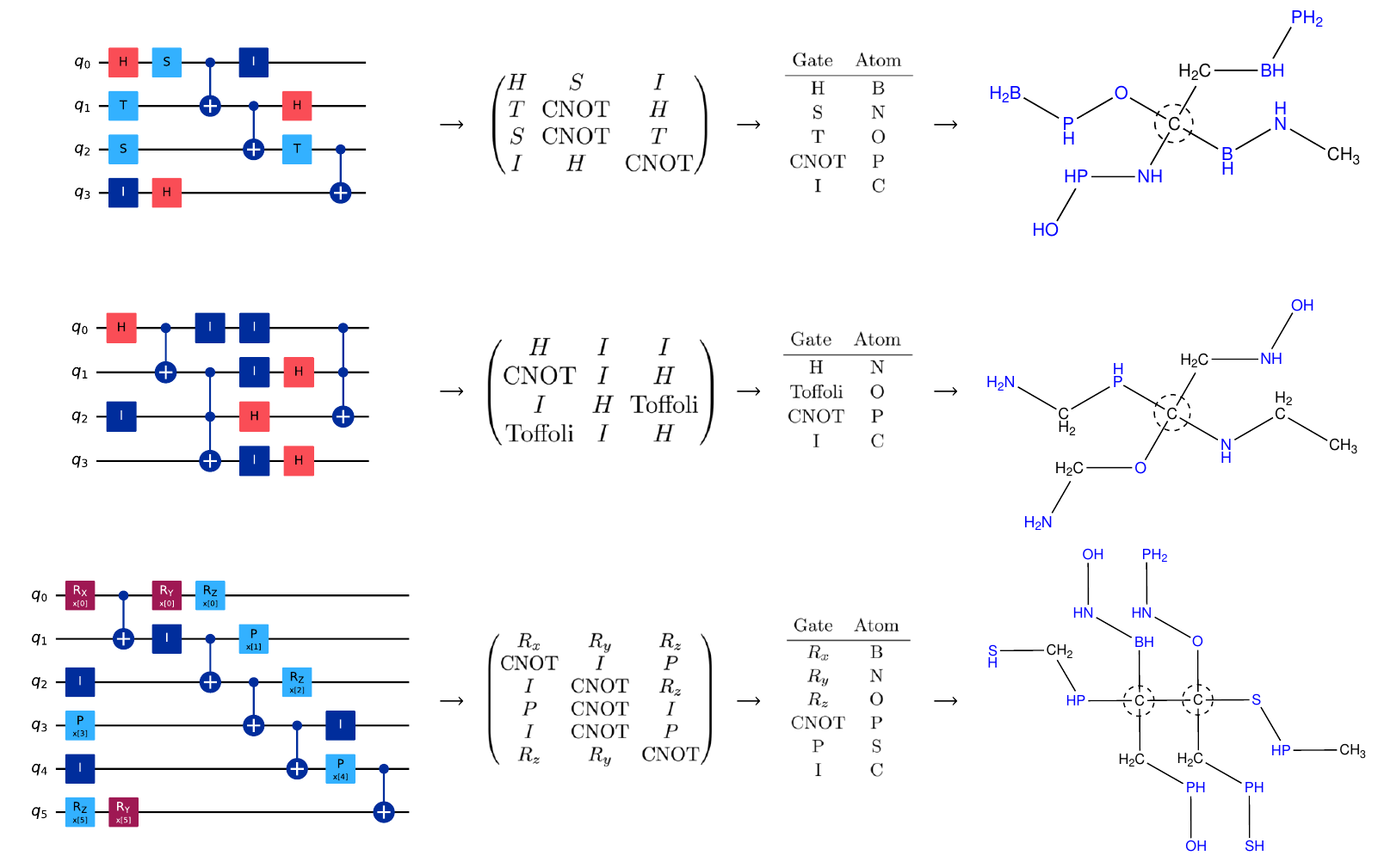}
    \caption{Three examples of mapping of quantum circuits with universal gate sets and the corresponding matrix representation of gates to molecules. Each qubit is represented by a molecular branch stemming from the backbone carbon atom/chain, encircled by the dashed line.}
    \label{fig:1}
\end{figure*}

\section{Isomorphism between QC and molecules}

We begin by establishing an isomorphism between universal QC and a specifically defined sub-space of polyatomic molecules. To prove isomorphism, we construct a mapping of QC with universal gate sets onto polyatomic molecules that ensures:
(1) bijection; (2) structure preservation — the mapping must preserve the essential relationships such as adjacency and order of operations, and (3) reversibility — there must exist an inverse mapping that is also structure-preserving.

A set of quantum gates is universal if the gates in this set can approximate any desired unitary operation to arbitrary precision \cite{kitaev2002classical}. 
This implies that a universal set of gates must support arbitrary single-qubit rotations and, at least, two-qubit entanglement \cite{nielsen2010quantum}.  
We consider, separately, three universal gate sets: set (a) including the rotation operators $R_x(\theta)$, $R_y(\theta)$, and $R_z(\theta)$, the phase shift gate $P(\phi)$ and CNOT  \cite{kitaev2002classical, Williams2011}; (b) the Clifford gates, including $\text{CNOT}$, the Hadamard gate $H$ and the phase gate $S$, supplemented by the phase-shift $T$ gate; (c) a combination of the Toffoli gates $T$ and the Hadamard gates \cite{Aharonov}.  
Though the Clifford set is not universal, as Clifford circuits can be simulated classically \cite{aaronson2004improved}, the addition of the $T$ gate makes the set universal. 
Similarly, the Toffoli gate alone can be used to simulate any classical reversible circuit, but is not sufficient for universal quantum computation. However, when combined with a quantum gate, such as $H$, the Toffoli gates can be used to build universal quantum circuits. 

We aim to obtain molecular representations that encode the architecture of universal QC applied to an arbitrary number $n_Q$ of qubits. The representations adopted here encode the type and positions of gates in the quantum circuits. We note that each QC can be uniquely represented by an $n_Q \times L$ matrix (shown in Figure 1), where $L$ is the number of gate layers. The matrix rows represent the qubits, and the columns specify the gates or no-gate (identity $I$), which are applied to each qubit in a given circuit layer. 
We assign each row of the matrix to a branch of atoms stemming from a central carbon atom chain
and identify each type of quantum gate with an atom that permits an appropriate number of covalent bonds.  
The identity operator is specified by the carbon atom, while the quantum gates are specified by non-carbon atoms placed in appropriate positions. 

We encode the number of qubits into a polymeric carbon chain as illustrated in Table I.
The goal is to ensure that the resulting molecules have $n_Q$ branches. For example, a 4-qubit circuit can be represented by a molecule with four branches of an $sp^3$-hybridized carbon atom, while a 6-qubit circuit requires two covalently bonded $sp$-hybridized carbon atoms. A molecule with an arbitrary number of branches can be designed by inserting the appropriate number of carbon atoms in the central carbon atom chain. We choose the hybridization of the carbon atoms in the backbone chain to represent $n_Q$ branches with the lowest number of carbon atoms. After assigning atoms to each molecular branch, we saturate the valency of all atoms by attaching hydrogen atoms to create stable molecules. A QC with an odd number of qubits is obtained by replacing one molecular branch with a hydrogen atom.

Each row of the matrix in Figure 1 is represented by a molecular branch with quantum gates described by specific atoms that permit, at least, two covalent bonds. 
Because the pool of such atoms is limited, we restrict the entanglement gates between qubits with $|i - j |\leq \delta$, where $i$ and $j$ are the row indices of the matrices in Figure 1 and $\delta$ is some finite integer. 
This does not restrict the universality of the resulting quantum circuits, as nearest-neighbour entanglement is sufficient to generate universal QC \cite{fellner2022universal,klaver2024swap}.  
Thus, we need 3 to 5 bi-valent atoms to identify all gates in any universal quantum gates set.

\begin{table}[H]
    \centering
    \renewcommand{\arraystretch}{2} 
    \begin{tabularx}{0.7\linewidth}{ c c }
        Number of qubits\;\; &  Carbon backbone \\
        \hline \hline
        4 &  \chemfig{C(-[:90,1])(-[:180,1])(-[:270,1])(-[:0,1])} \\[2em] 
        6 &  \chemfig{C(-[:90,1])(-[:180,1])(-[:270,1])(-[:0,1])-[::+0,1.3]@{=}C(-[:90,1])(-[:180,1])(-[:270,1])(-[:0, 1])} \\[2em]
        8 &  \chemfig{C(-[:90,1])(-[:180,1])(-[:270,1])(-[:0,1])-[::0,1.3]C(-[:90,1])(-[:270,1])(-[:0,1])-[::0,1.3]C(-[:90,1])(-[:180,1])(-[:270,1])(-[:0,1])} \\[2em]
        \vdots & \vdots \\[0.5em]
        $n_Q$ & $n_{\rm C} = \frac{n_Q}{2} -1$ \\ \hline \hline
    \end{tabularx}
    \caption{Central carbon atom chain designed to produce $n_Q$ covalently bonded branches.  As illustrated, the carbon polymer chain requires $n_{\rm C}$ carbon atoms for even $n_{\rm Q}$. A QC with an odd number of qubits can be obtained by replacing one molecular branch with a hydrogen atom.}
    \label{tab:1}
\end{table}

\section{Methodology for numerical calculations}

\subsection{Molecular descriptors}
\label{molecular_descriptors}

For machine learning and cheminformatics applications, molecules are encoded into molecular descriptors. 
Molecular descriptors must generally be numerical, universal, and capable of capturing essential features of molecules such as molecular shape, chemical bond arrangements, bond strengths and orders, vibrational frequencies, and conformational flexibility. 
While it is possible to design unique multi-dimensional descriptors, such as molecular fingerprints \cite{rogers2010extended} or graph neural networks (GNNs) \cite{kipf2016semi,wu2020comprehensive} to explicitly represent the spatial arrangement and connectivity of atoms within a molecule, much recent work has focused on the development of physical descriptors, including 
Coulomb matrices \cite{rupp2012fast}, bag of bonds (BoB) features \cite{hansen2015machine}, smooth overlap of atomic positions (SOAP) \cite{bartok2013representing}, and the Faber–Christensen–Huang–Lilienfeld (FCHL19) representation \cite{christensen2020fchl}. 
Molecular fingerprints can be viewed as a digital signature of molecules, encapsulating the structural and chemical attributes through strings with 1000 to 4000 bits. Such descriptors are useful for cheminformatics applications, but impractical for interpreting the effects of individual atom groups on molecular properties or for optimization of molecular properties in chemical compound spaces. On the other hand, physical descriptors are more suitable for algorithms aiming to optimize the molecular properties, e.g., in spaces of molecular isomers \cite{mao2024efficient}. 

To demonstrate the feasibility of molecular representations of quantum circuits, we first consider molecular fingerprints as descriptors of QC. 
We use count-based molecular fingerprints implemented in RDKit and inspired by the Daylight fingerprinting method \cite{Daylight}. These fingerprints identify subgraphs of a molecule within a specified range of molecular groups, hash them into bit IDs, and use modulo operations to map these IDs to a fingerprint of fixed size. The count-based format records the frequency of each subgraph occurrence rather than a binary presence/absence. The hashing process incorporates atomic and bond properties, such as the atomic number (modulo 128), aromaticity, atom degree, and bond type \cite{RDKit}.

In order to explore practical applications of molecular descriptors for structural optimization of quantum circuits, we also employ the Coulomb matrix representations of molecules. 
As shown by Rupp et al. \cite{rupp2012fast}, molecules can be effectively described by a symmetric matrix with the matrix elements given by
\begin{equation}
    M_{ij} =
    \begin{cases}
        0.5 Z_i^{2.4} & \quad \text{for}~i=j\\
        Z_i Z_j/r_{ij} & \quad \text{for}~i \neq j,
    \end{cases}
\end{equation}
where $Z_i$ and $Z_j$ are the atomic numbers of atoms $i$ and $j$, respectively, $r_{ij}$ is the distance between atoms $i$ and $j$, with the indices $i,j \in [1, n_{\rm max}]$, and $n_{\rm max}$ is the number of atoms in the molecule. 
The diagonal elements of the Coulomb matrix represent a polynomial fit relating the atomic number to the total energy of free atoms \cite{rupp2012fast}, while the off-diagonal elements describe the electrostatic interaction between atoms $i$ and $j$. 
The Coulomb matrix is invariant to molecular translations and rotations, though not invariant to atomic permutations.

It was previously shown that the eigenvalues of Coulomb matrices contain physical information about molecules \cite{schrier2020can}. However, within the QC-molecule mapping scheme proposed here, the number of atoms grows polynomially with the number of qubits and gates in the corresponding QC. 
Therefore, it is impractical to consider full Coulomb matrices or their eigenvalues as descriptors of QC. Instead, we propose and explore the application of 
the Gershgorin circle theorem. This theorem identifies a region in the complex plane that includes all eigenvalues of a complex square matrix \cite{schrier2020can}. More specifically, 
for an $m \times m$ matrix with complex entries, the eigenvalues of matrix $M$ lie within the union of discs $D_1 \cup D_2 \cup \dots \cup D_m$, where each disc $D_i$ is defined as
\begin{equation}
    D_i = \left\{ z \in C : | z - M_{ii} | < \sum_{j\neq i} |M_{ij}|\right\}.
    \label{gershgorin}
\end{equation}
 
We use the radii of the smallest and largest Gershgorin circles for the Coulomb matrices of the corresponding molecules to characterize the QC and to restrict the structural optimization of QC. 

To generate molecular structures, we use RDKit \cite{RDKit}. More specifically, for the calculations presented here, we use molecular structures with the three-dimensional geometry optimized by the universal force field (UFF) method. The UFF optimization minimizes potential energy, including van der Waals interactions, and accounts for torsional interactions, in addition to energies of chemical bonds. This yields physical sterical conformations \cite{rappe1992uff}. We have also repeated the present calculations with molecules represented by two-dimensional arrangements of atoms. The 2D coordinates are computed using the Kamada-Kawai force-directed algorithm, which optimizes atomic positions by minimizing an energy function based on linear springs \cite{kamada1989algorithm}. This method preserves atomic connectivity but does not account for steric effects. Since UFF optimization is computationally expensive and scales significantly with the increasing number of atoms, bonded and non-bonded interactions \cite{jasz2019optimized}, using 2D representations reduces the computational cost of generating molecules. Furthermore, a 2D molecular representation can be directly mapped to the coordinates of neutral atoms on quantum platforms with well-defined and well-controlled 2D atom arrays, such as Pasqal \cite{scholl2021quantum} and Quera \cite{ebadi2022quantum}, which allows the application of particular quantum algorithms for determining molecular properties on such quantum computers \cite{d2024leveraging,maskara2025programmable}.

\subsection{Quantum support vector machines}

\label{qsvm}

In the present work, we consider applications of QC for quantum kernels of support vector machines (SVM). The resulting SVMs are used for binary classification problems. The inputs $\bm x_i \in \mathbb{R}^n$ of the dataset ${\cal D} : = \{\bm x_i, y_i \in [0,1] \}_{i=1}^N$ are encoded into quantum kernels 
\begin{align}
k(\bm x, \bm x') =    | \bra{\Phi ({\bm x'})}  {\cal U}^\dagger(\bm x') {\cal U}(\bm x)  \ket{\Phi ({\bm x})} |^2,
\label{q-kernels}
\end{align}
where
\begin{align}
    \ket{\Phi ({\bm x})}= \exp{\left( i\displaystyle \sum_{k} x_k Z_k \right)} H^{\bigotimes n} \ket{0}^{\bigotimes n},
    \label{map}
\end{align}
$x_k$ is the $k$-th component of $\bm x$, $Z_k$ is the Pauli $Z$-gate acting on $k$-th qubit, and $H$ is the Hadamard gate.
The sequence of gates in ${\cal U}(\bm x)$ determines the performance of the quantum SVM models \cite{torabian2023compositional}. We choose $\cal U$ to consist of $L$ layers, each including one-qubit $R_Z$ gates and/or two-qubit CNOT gates. Each of the $R_Z$ gates for qubit $k$ is parametrized by $\theta$ as follows: 
\req{
R_{Z,k}(\theta)
=
  \begin{pmatrix}
 e^{-i \theta x_k} &
   0 \\
   0 &
   e^{i \theta x_k} 
   \end{pmatrix}.
 }
This yields a quantum circuit with the number of free parameters $\theta$ equal to the number of $R_Z$ gates. We optimize the parameters of all generated QCs using Bayesian optimization (BO) \cite{shahriari2015taking}.

To illustrate that molecular descriptors can be used as representations of quantum circuits, we consider four distinct classification problems: (I) given the ionic radii of A, B, B$^\prime$ and X, classification of halide perovskites with the chemical formula A$_2$BB$^{\prime}$ X$_6$ into metals or non-metals \cite{jain2013commentary} ; (II) classification of a four-dimensional (4D) synthetic data set implemented in \cite{bowles2024better}; (III) classification of a five-dimensional (5D) synthetic data set implemented in \cite{bowles2024better}; (IV) five-dimensional classification of digits 3 and 5 based on the MNIST dataset from Ref. \cite{lecun1998mnist}.
Hereafter, we refer to these classification problems as perovskite (for problem I), hidden-manifold (for problems II and III), and MNIST (for problem IV). For each classification problem, we generate $N$ QC ${\cal U}(\bm x)$ including $L$ layers of randomly sampled quantum gates, and train QSVM with the corresponding quantum kernels. The accuracy of the resulting QSVM is used as a metric of the corresponding quantum kernel performance. We use problems I and II to identify the molecular properties that can be used to optimize the search of performant QC. We then use problems III and IV to demonstrate the performance of the algorithm thus developed.

We use the following metric to quantify the performance of a quantum kernel (of equivalently a QC). 
The dataset for each classification problem is split into a training set and a test set. The training set for the perovskite classification problem includes 100 data points. 
For all other classification problems considered in this work, the training set includes 1000 data points. 
The test sets comprise 1442 randomly sampled materials for the perovskite classification problem and 1000 randomly selected data points for all other classification problems considered in this work.  
We label the two classes of each classification problem as positive and negative and define the true positive rate (TPR) and true negative rate (TNR) as the ratio of correct predictions to the total number of test points for a given class. 
The average accuracy of an SVM model is defined here as the balanced average of TPR and TNR computed over the test set.
The boundary separating performant and underperforming circuits is defined as the midpoint between the maximum and minimum average classification accuracy over the entire set of quantum circuits. 
We include a margin of $\pm 10~\%$ to define a QC as performant, when the average accuracy of the corresponding SVM exceeds the boundary by more than 10 \%, and underperforming when the resulting average accuracy is more than 10 \% below the separating boundary. 
We discard the QC within the $\pm 10~\%$ margin of the separating boundary.

\subsection{Simplified QC - molecule mapping}
\label{simplified-mapping}

As shown in Section II, the operators ${\cal U}(\bm x)$ can be built with the universal set of quantum gates. In order to make computations more efficient, for numerical examples, we use two simplifications: (1) we allow ${\cal U}(\bm x)$ to include only one-qubit $R_Z$ gates and two-qubit CNOT gates; (2) we omit the identity operators from the QC $\rightarrow$ molecule mapping. 
 This makes the inverse mapping from molecules to QC non-unique. 
 However, our numerical tests show that this does not affect the results presented in this work.

The following section considers numerical examples with 4-qubit  and 5-qubit  QC. 
To map quantum gates to atoms for the numerical examples, we utilize the following scheme: $R_Z \rightarrow \mathrm{C}$, $\text{CNOT}(\delta = 1) \rightarrow  \mathrm{N} $, $\text{CNOT}(\delta = 2) \rightarrow \mathrm{O} $, $\text{CNOT}(\delta = 3) \rightarrow \mathrm{S}$, and $\text{CNOT}(\delta = 4) \rightarrow \mathrm{P}$.
The carbon atom backbone is constructed as illustrated in Table I.   In particular, for the 5-qubit circuits, we use two $sp^3$-hybridized atoms with the valency of one of these atoms saturated by one hydrogen atom. 

We consider two types of problems. First, we examine the performance of QC for QSVM in the space of molecular descriptors in order to show that molecular descriptors can serve as meaningful descriptors of QC. 
Second, we build molecules by constraining the algorithm to particular parts of a molecular space, and transform these molecules into QC to illustrate that molecular descriptors can reduce the complexity of compositional QC optimization. 

\begin{figure*}[t!]
    \centering
    \includegraphics[width=0.9\linewidth]{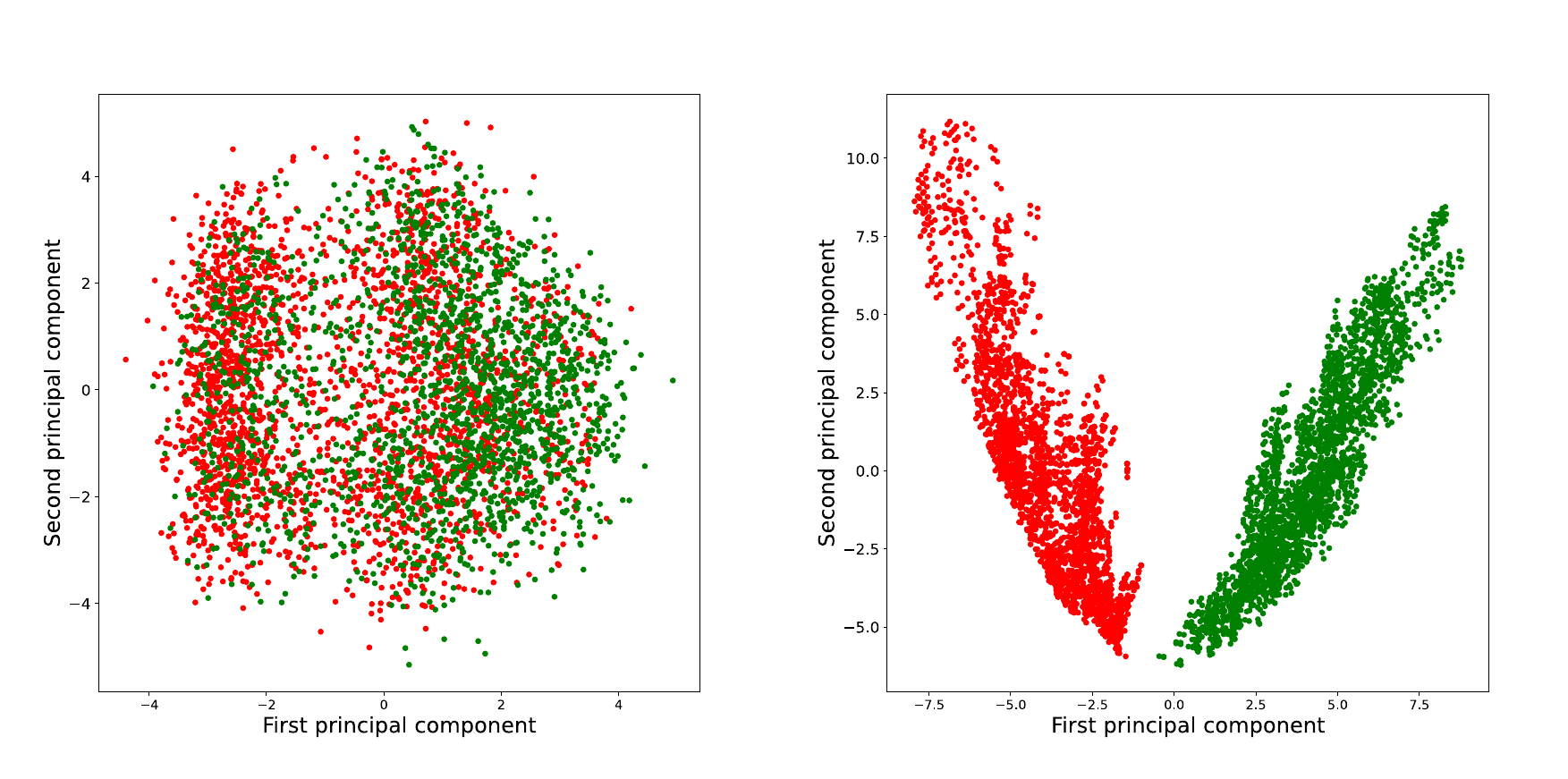}
    \caption{Distributions of performant QC (red circles), yielding high classification accuracy of QSVM, and underperforming QC (green triangles), leading to low classification accuracy of QSVM, in the space of two leading principle components after PCA dimensionality reduction of molecular fingerprints. 
    The results are obtained with 10000 randomly generated QC used to build kernels for QSVM to classify the 4D perovskite (left panel) and the 4D hidden-manifold (right panel) classification problems. 
    }
    \label{fig:2}
\end{figure*}

    \begin{figure*}
        \centering
        \includegraphics[width=1.05\textwidth]{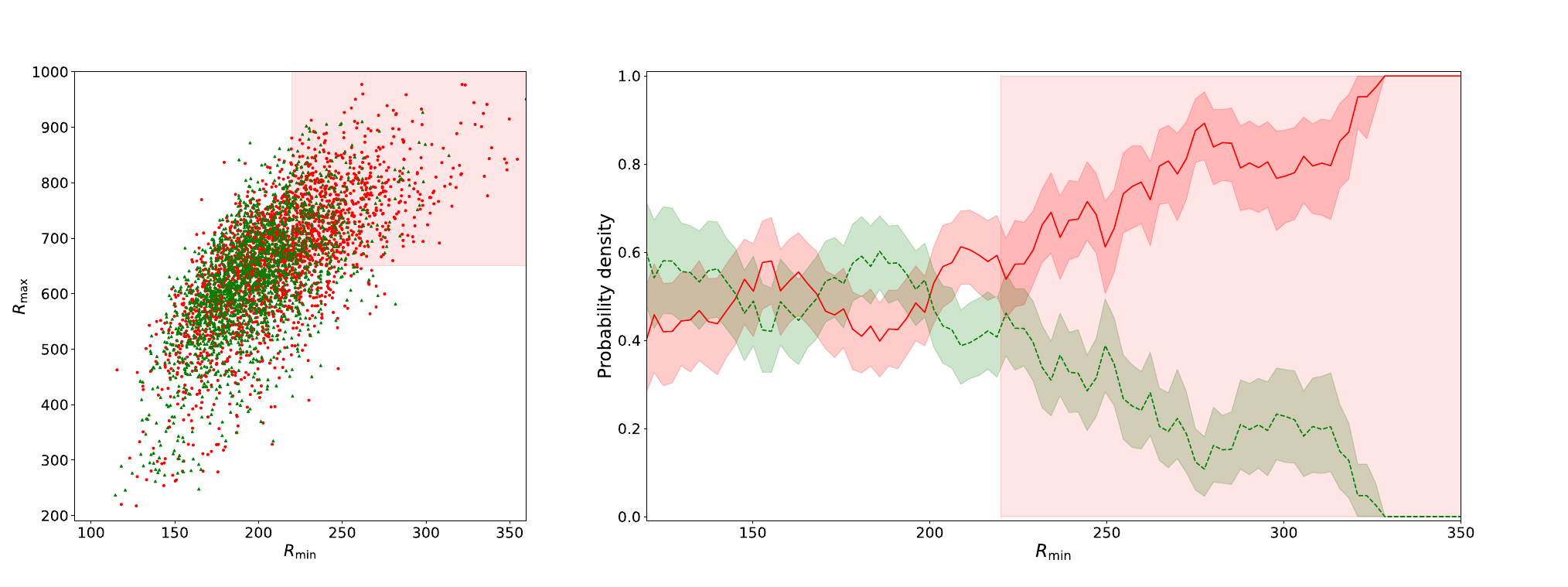}\\
        \includegraphics[width=1.05\textwidth]{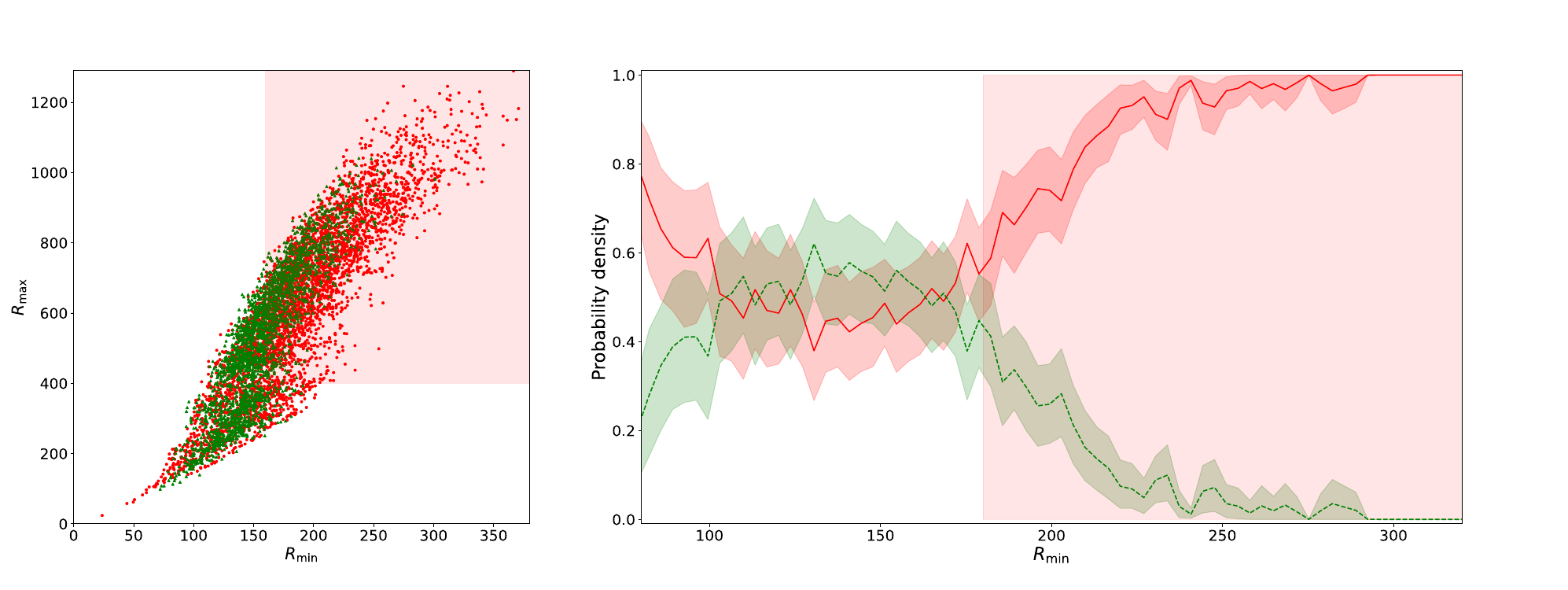}
        \caption {Left panels: Distributions of performant QC (red circles) and underperforming QC (green triangles) in the space of the smallest ($R_{\rm min}$) and largest ($R_{\rm max}$) Gershgorin radii for the 4D perovskite (top) and 4D hidden-manifold (bottom) classification problems. 
        Right panels: Probability density of performant (solid red curves) and underperforming (broken green curves) as functions of $R_{\rm min}$ computed by KDE from the data in the left panels. The shaded region indicates the $95\%$ confidence interval, reflecting the uncertainty in the density estimation due to the finite sample size.
      } 
        \label{fig:3}
    \end{figure*}

    \begin{figure*}
        \centering
        \includegraphics[width=0.48\linewidth]{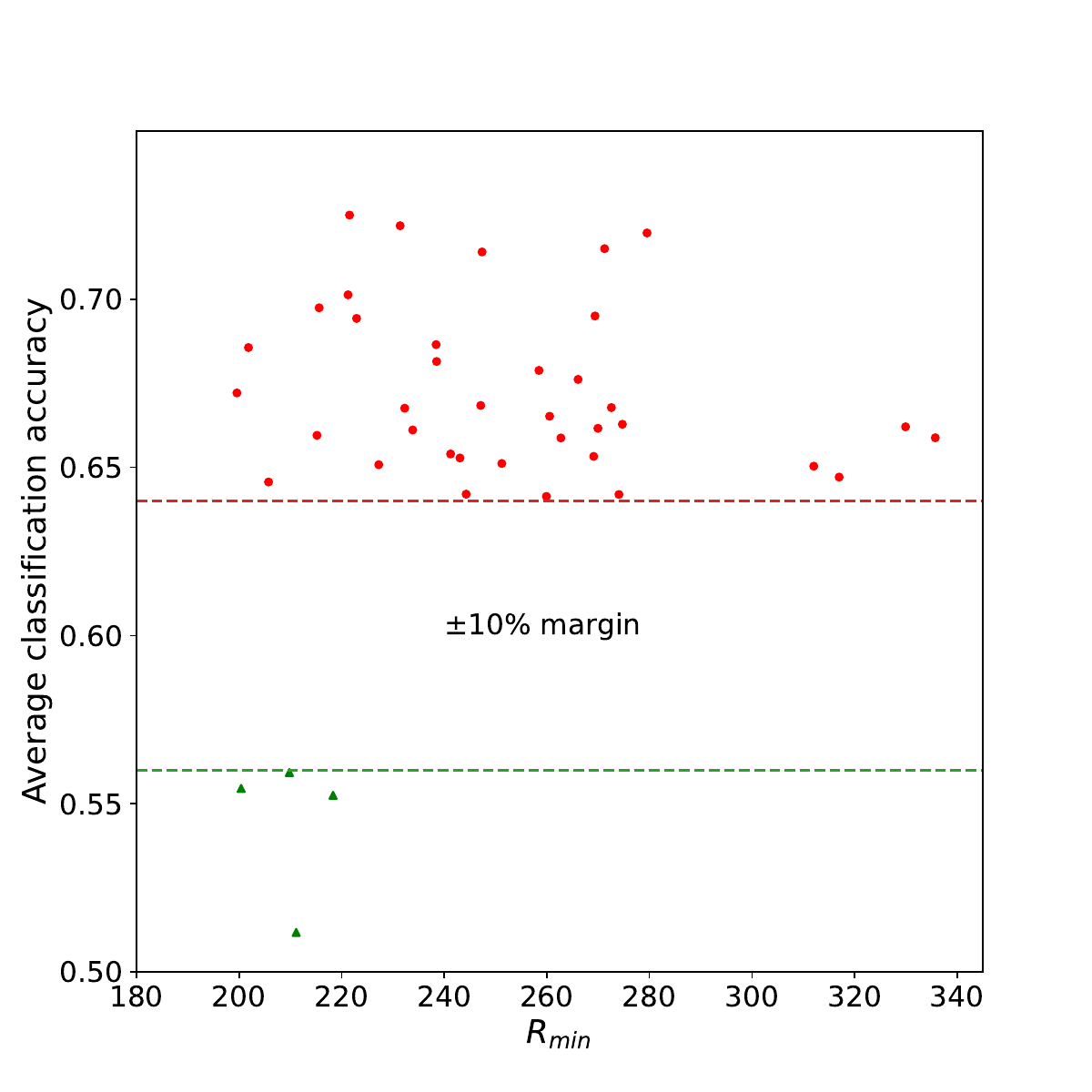}
        \includegraphics[width=0.48\linewidth]{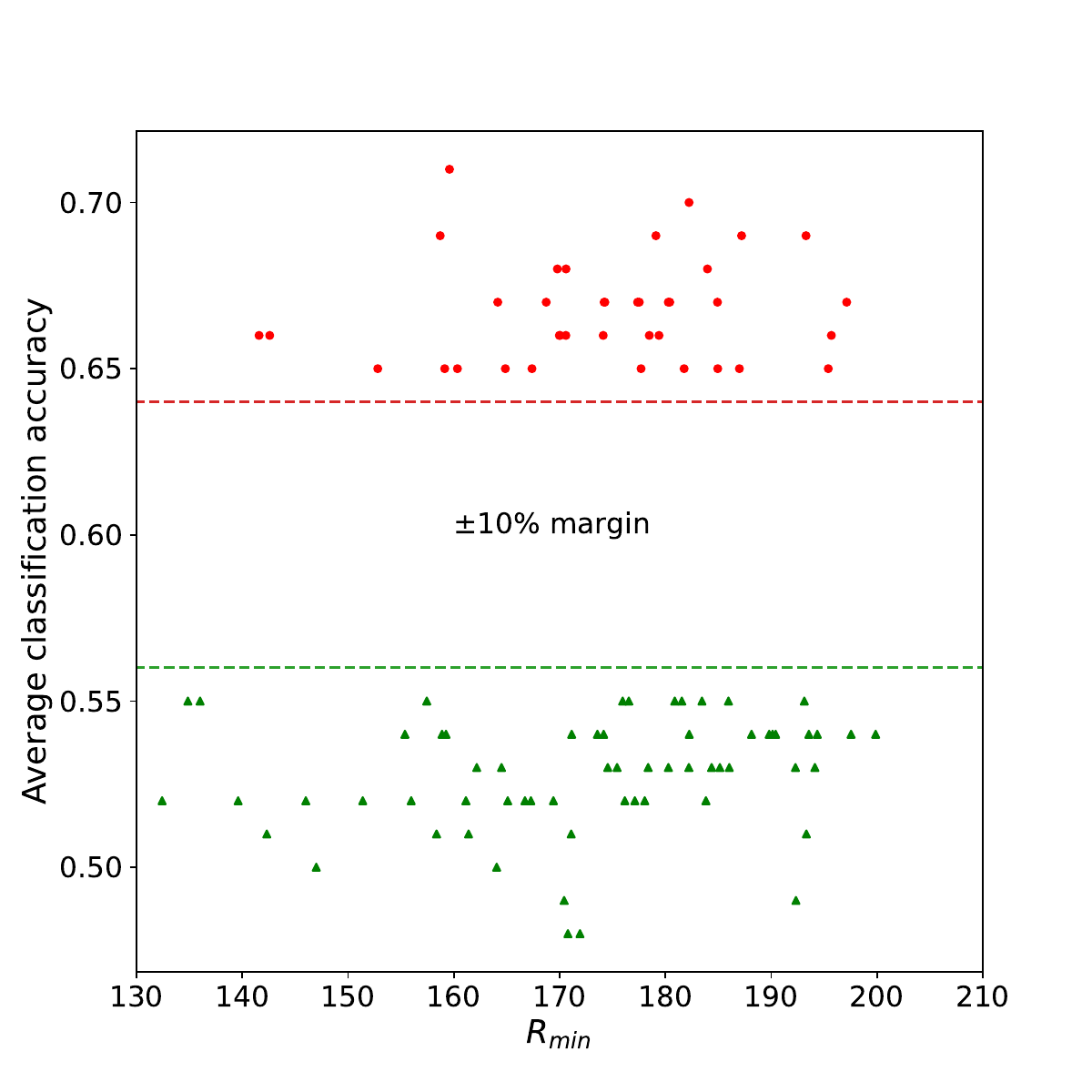}
        \caption{Distributions of average QSVM accuracy for the 4D perovskite classification problem with five-layer QC constructed from 100 randomly generated molecules with large radii ($R_{\rm min} >220$) of the smallest Gershgorin circle (left panel) and low radii ($R_{\rm min} < 220$) of the smallest Gershgorin circle (right panel). The dashed lines show the $\pm 10\%$ margin centered at the average of the minimum and maximum classification accuracy in a sample of 10000 random quantum circuits. The red circles and green triangles represent performant and underperforming QC designs, respectively.}
        \label{fig:4}
    \end{figure*}

    \begin{figure*}
        \centering
        \includegraphics[width=0.48\linewidth]{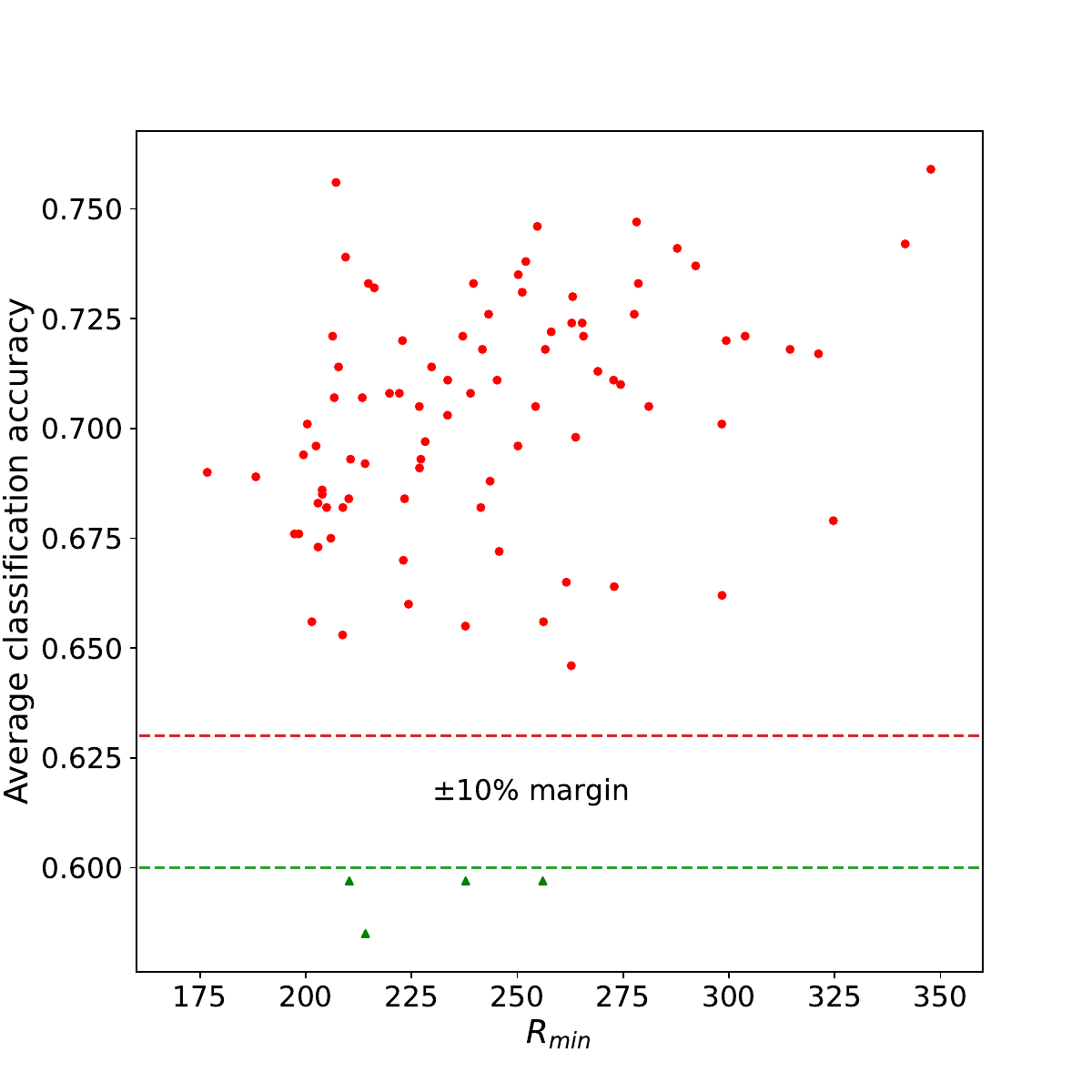}
        \includegraphics[width=0.48\linewidth]{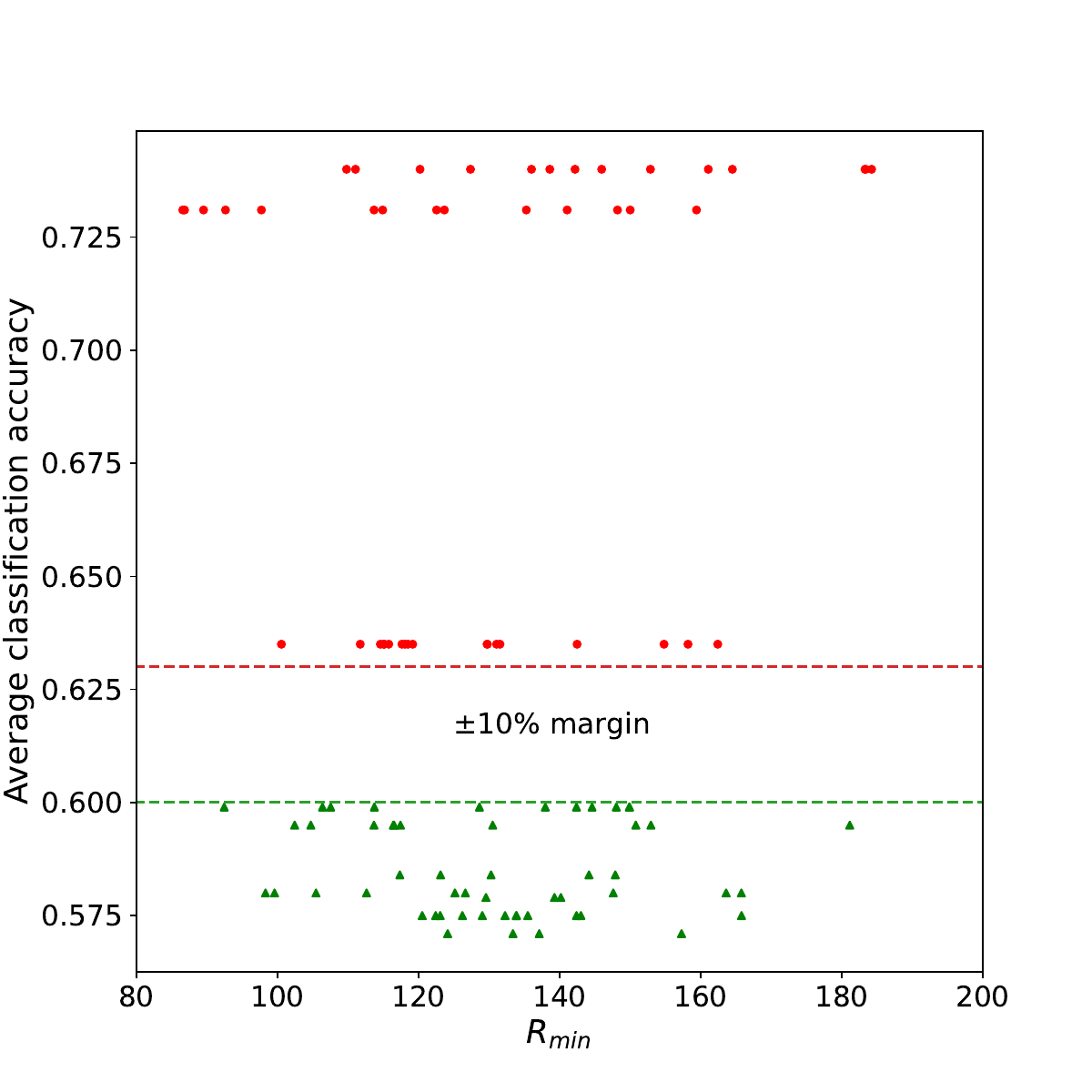}
        \caption{Distributions of average QSVM accuracy for the 4D hidden-manifold classification problem with five-layer QC constructed from 100 randomly generated molecules with large radii ($R_{\rm min} >170$) of the smallest Gershgorin circle (left panel) and low radii ($R_{\rm min} < 170$) of the smallest Gershgorin circle (right panel). The dashed lines show the $\pm 10\%$ margin centered at the average of the minimum and maximum classification accuracy in a sample of 10000 random quantum circuits. The red circles and green triangles represent performant and underperforming QC designs, respectively.}
        \label{fig:5}
    \end{figure*}

    \begin{figure*}
        \centering
        \includegraphics[width=0.95\linewidth]{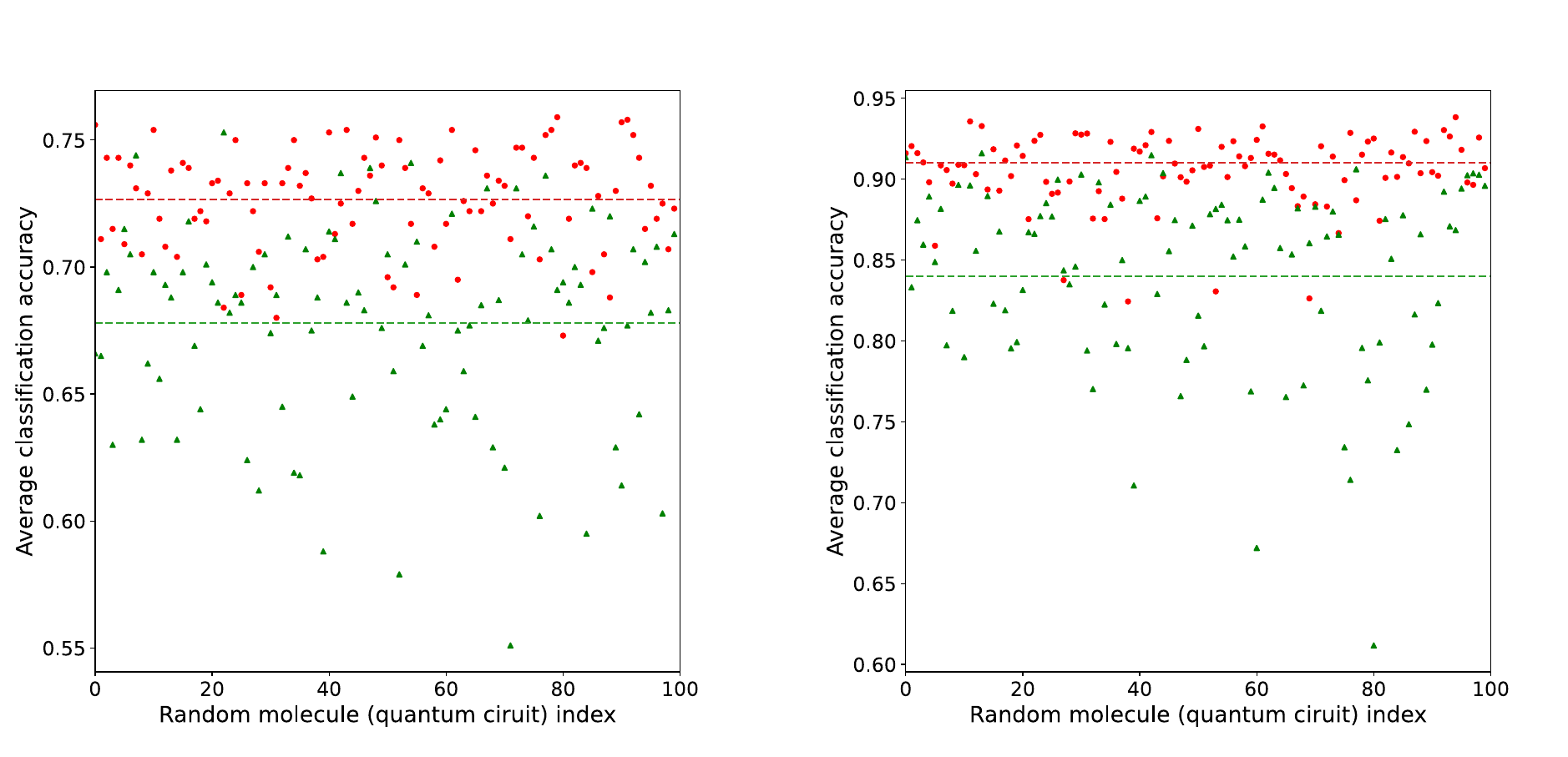}
        \caption{
        Distribution of average QSVM accuracy for the 5D hidden-manifold (left) and 5D MNIST (right) classification problems using five-layered QC obtained from 10000 randomly generated molecules. The red circles represent a subset of 100 molecules with the largest radius of the smallest Gershgorin circle and the green triangles represent a subset of 100 molecules with the smallest radius of the smallest Gershgorin circle. The dashed lines show the $\pm 10\%$ margin centered at the average of the minimum and maximum classification accuracy in the entire sample of 10000 quantum circuits. }
        \label{fig:6}
    \end{figure*}

    \begin{figure*}
        \centering
        \includegraphics[width=1.05\textwidth]{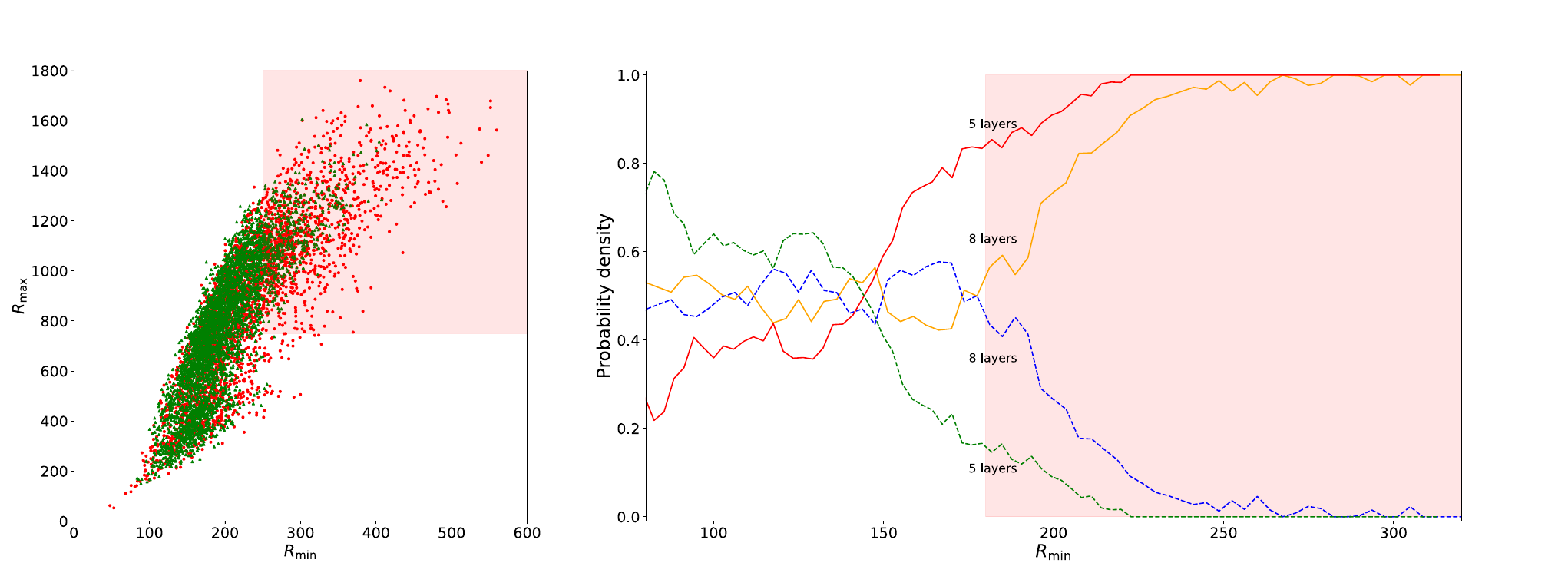}
        \caption{Left: same as in Fig. \ref{fig:3} (lower left) but for QC with eight layers. 
        Right: Probability density of performant (solid curves) and underperforming (broken curves) QC as functions of $R_{\rm min}$ computed for QC with five and eight layers.}
        \label{fig:7}
    \end{figure*}

\section{Results}

\subsection{Molecular fingerprints as QC descriptors}

We begin by illustrating the feasibility of applying widely used molecular fingerprints as variables to characterize the performance of QC for quantum SVM. 
For this calculation, we generate $N=10,000$ QC ${\cal U}(\bm x)$ with $L=5$ layers of randomly sampled quantum gates and train QSVM with the corresponding quantum kernels.
Each QC is mapped onto a molecule, as described in Section \ref{simplified-mapping}. High-dimensional fingerprints are constructed for the corresponding molecule using the method in Ref.  \cite{Daylight} as described in section \ref{molecular_descriptors}. 
The dimensionality of the molecular descriptors is then reduced by principal component analysis (PCA) to two principal components. Each of principal components is a linear combination of fingerprints components.

Figure \ref{fig:2} shows the performance of QC for QSVM applied to two classification problems described in Section \ref{qsvm} as a function of the two principle components thus obtained. 
Here and hereafter, the QSVM accuracy is represented by red circles for performant circuits and green triangles for underperforming kernels, yielding low accuracy of QSVM.
The distinction between the performant and underperforming QC adopted here is described in Section \ref{qsvm}.
Figure \ref{fig:2} demonstrates that there is a separation between performant QC and underperforming QC in the PCA-reduced space of molecular fingerprints. 
This separation is particularly clear for the 4D hidden-manifold classification problem, in which the data for two classes are generated by a classical artificial neural network. 
This suggests that PCA-reduced molecular fingerprints can be used to restrict the search of performant QC for QSVM through the QC $\leftrightarrow$ molecule mapping algorithm proposed here.

While Figure \ref{fig:2} establishes a meaningful correlation between molecular descriptors and QC performance, reducing the dimensionality of molecular fingerprints by PCA requires diagonalization of high-dimensional matrices. The computational complexity of matrix diagonalization
scales as $N_{\rm dim}^3$ with the matrix dimensionality $N_{\rm dim}$, making PCA-reduced molecular fingerprints computationally expensive for large QC.  Furthermore, the complexity of molecular fingerprints complicates the interpretability of the basic elements 
that make a QC performant. Therefore, the following section explores alternative molecular representations of QC.

\subsection{Gershgorin circles of Coulomb matrices as QC descriptors} 
    
\label{gershgorin-section}

A key goal of the present work is to identify molecular properties that are (i) either scalar or low-dimensional vectors; (ii) sensitive to the structural composition and shape of polyatomic molecules; (iii) efficient to compute, and (iv) provide a meaningful characterization of the corresponding QC performance.     
To achieve this, we explore the correlation between the performance of QC for quantum SVM and the values of the radii of the Gershgorin circles (\ref{gershgorin}) derived from the Coulomb matrices for the corresponding molecules. 
The Gershgorin circle radii are represented by sums of the matrix elements and can thus be efficiently computed. 
For each molecule, we consider the radii of the largest and smallest Gershgorin circles, denoted for a given Coulomb matrix by $R_{\rm min}$ and $R_{\rm max}$ respectively.

The left panels of Figure \ref{fig:3} illustrate the QC performance in the two-dimensional space of $R_{\rm min}$ and $R_{\rm max}$. The right panels of Figure \ref{fig:3} show the probability density of performant and low-accuracy QC computed from the distributions in the left panels by kernel density estimation (KDE) as functions of $R_{\rm min}$.
The results are presented for the 4D perovskite (upper panels) and 4D hidden-manifold  (lower panels) classification problems. Figure \ref{fig:3} shows that the magnitudes of the smallest and largest Gershgorin circles for a given molecule can be used to guide the search of performant QC for quantum SVM. To illustrate this directly, we generate two sets of 100 random molecules: one sampling from the quadrant with $R_{\rm min} > 220$  and $R_{\rm max} > 650$ (higher probability of finding performant QC), and another sampling from the quadrant with $R_{\rm min} < 220$ and $R_{\rm max} < 650$ (lower probability of finding performant QC). 

All molecules in these random sets correspond to QC with 5 layers. We map each of these molecules onto a quantum kernel and apply the kernel thus obtained to the perovskite classification problem. The accuracy of the resulting models is shown in Figure \ref{fig:4}, with the results obtained from molecules with $R_{\rm min} > 220$ shown in the left panel and the results from molecules with $R_{\rm min} < 220$ in the right panel. The horizontal dashed line represents the average accuracy of 10,000 QSVM models illustrated in Fig.  \ref{fig:3}.  
Figure \ref{fig:5} presents the same results but for the 4D hidden manifold dataset and with the panels corresponding to sampling from the quadrant with $R_{\rm min} > 170$ and $R_{\rm max} > 400$ and the quadrant with $R_{\rm min} < 170$ and $R_{\rm max} < 400$.

Note that the random samples of molecules used in  Figs. \ref{fig:4} and \ref{fig:5} are independent of the set in Fig.  \ref{fig:3}. The left panels of Figs. \ref{fig:4} and \ref{fig:5} show that randomly sampled molecules with a large value of $R_{\rm min}$ are much more likely to produce a performant quantum kernel for both datasets (e.g. the number of red circles is much larger than the number of green triangles). The right panels of Figs. \ref{fig:4} and \ref{fig:5} illustrate that the opposite 
is also true: randomly sampled molecules with a small value of $R_{\rm min}$ are much more likely to produce quantum kernels yielding low-accuracy models.

We next illustrate the proposed algorithm of designing performant QC by engineering polyatomic molecules for two independent classification problems: a five-dimensional version of the hidden-manifold classification \cite{bowles2024better} and the five-dimensional classification of digits 3 and 5 based on the MNIST dataset from Ref. \cite{lecun1998mnist}. 
We begin by generating a random sample of 10,000 molecules, each corresponding to QC with five layers of gates, for each classification problem. We then map 100 molecules with the highest values of the smallest Gershgorin circle radius and 100 molecules with the lowest values of the smallest Gershgorin circle radius onto the corresponding quantum circuits. Using QC thus constructed, we build quantum SVM models for the 5D hidden-manifold and 5D MNIST classification problems. The results are presented in Fig. \ref{fig:6}.

Designing quantum kernels by molecule $\rightarrow$ QC mapping provides insights into what determines the performance of quantum kernels for QSVM. 
Formally, a large radius of a Gershgorin circle indicates that, for the corresponding row of the matrix, the sum of the magnitudes of the off-diagonal elements is large compared to the magnitude of the diagonal element. 
For Hamiltonian matrices, a large Gershgorin circle may indicate strong coupling between different states, where the results are less determined by the local properties (diagonal elements) than by interactions with other states (off-diagonal elements). 
For molecules represented by Coulomb matrices, a large radius of the smallest Gershgorin circle suggests that the corresponding atom (associated with that particular row of the matrix) interacts strongly with many other atoms in the molecule. 
This implies a compact shape of tightly packed atoms. A large value of $R_{\rm min}$ also suggests that the central atom -- one of the carbons in the backbone chain for the scheme adopted here --  is in a highly symmetric or well-coordinated environment. 
For example, highly symmetrical molecules such as benzene or fullerenes include atoms that have strong and relatively uniform interactions with multiple neighbouring atoms, leading to large radii of the Gershgorin circles. 
For quantum kernels produced by the molecule $\rightarrow$ QC mapping, this implies that the performance of QSVM is enhanced by increasing the density of the gate layout.

A second key goal of this work is to develop an algorithm for identifying QC performance metrics based on a set of shallow QC in order to design deep QC that yield high accuracy in the corresponding quantum model. 
The QC $\leftrightarrow$ molecule mapping is well suited for this purpose as molecules can be grown incrementally by increasing the length of the molecular branches. The shape of polyatomic molecules thus designed is determined by the shape of the parent molecules by construction. 
To confirm this by numerical examples, we extend the calculations in Fig. \ref{fig:3} (lower panels) to QC with additional layers of quantum gates. The results of  Fig. \ref{fig:3}  are based on QC with five layers of gates, whereas Fig. \ref{fig:7} depicts the distribution of QC performance in the ($R_{\rm min}, R_{\rm max}$) space for QC with eight layers.  
Fig. \ref{fig:7} shows that the performance of deeper QC is enhanced in similar parts of the molecular parameter space. 
The comparison of Figs. \ref{fig:3} and \ref{fig:7} suggests the following approach: first, a distribution of efficient QC with a small number of gate layers is used to identify the range of $R_{\rm min}$ and $R_{\rm max}$ associated with a high performance of the quantum algorithm. This range of 
$R_{\rm min}$ and $R_{\rm max}$ can then be used to restrict the search of deeper QC that yield high performance.

We have repeated the calculations in Figs. \ref{fig:1} -- \ref{fig:7} using 2D molecular representations. Our numerical tests indicate that using these representations produces similar distributions of performant and underperforming quantum circuits in the space of Gershgorin circle descriptors for QCs with five and eight layers, and for four- and five-dimensional datasets. Since optimization of 3D molecular structures is computationally more time-consuming, these results indicate that 2D representations can be used as a less expensive yet accurate alternative for mapping QCs to molecules. Furthermore, these results suggest the possibility of determining molecular properties using quantum platforms better suited for 2D graphs \cite{scholl2021quantum,ebadi2022quantum} as an alternative to experimental measurements or classical computations.

\section{Conclusion}

We have demonstrated the isomorphism between quantum circuits and a subspace of polyatomic molecules,
which suggests that molecules can be used as descriptors of quantum circuits. This connection has several important implications. First, the numerical techniques for optimization of molecular properties in chemical compound spaces can be adopted for optimization of quantum circuits for quantum algorithms. Thus, quantum ansatz optimization can benefit from advances in cheminformatics. In addition, a large number of descriptors developed for machine learning in chemistry can be applied to machine learning in quantum circuit spaces. In this work, we consider two qualitatively different molecular descriptors to characterize the performance of quantum circuits. 

We have shown that the performance of QC for QSVM can be characterized by PCA-reduced molecular fingerprints as well as by the size of the largest and smallest Gershgorin circles derived from the Coulomb matrices of the corresponding molecules. This can be used to restrict the search space for the compositional optimization of quantum circuits, as illustrated by Fig. \ref{fig:6} for two independent classification problems. We have shown that a high accuracy of a quantum algorithm can be achieved with high probability by sampling from a particular set of molecules.

The specific QC $\leftrightarrow$ molecule mapping proposed here provides an algorithm for increasing the depth of quantum circuits, while improving the QC performance. 
Since qubits are encoded by different branches from a carbon polymer chain, large polyatomic molecules representing deep quantum circuits inherit some properties of smaller molecules, corresponding to shallower quantum circuits.  This can be used to identify the range of molecular parameters yielding enhanced performance of a quantum algorithm with shallow circuits. The range of the molecular parameters thus determined can then be exploited to restrict the search of optimal quantum circuits of larger depth.

The Gershgoring circles of Coulomb matrices effectively encode information about the distribution of atoms as determined by the molecular shape.  Our results thus indicate that the shape of molecules used to construct quantum circuits is important for the performance of the resulting quantum algorithm. More broadly, the present work suggests an approach to identify key elements that enhance the accuracy of a quantum algorithm by mapping QC onto molecules and exploring correlations between physical properties or atomic compositions of molecules and circuit performance.

\section*{Acknowledgments}

This work was supported by NSERC of Canada. 

\clearpage

\bibliography{References.bib}

\end{document}